\documentclass[aps,prl,a4paper,preprintnumbers,amsmath,amssymb,superscriptaddress,twocolumn,floatfix,showpacs]{revtex4}
\usepackage{amssymb}
\usepackage{float}
\usepackage[pdftex]{graphicx}

\newcommand{\dd}{\,{\rm d}}

\newcommand{\GeV}{\,{\rm GeV}}

\newcommand{\kpc}{\,{\rm kpc}}

\newcommand{\keV}{\,{\rm keV}}
\newcommand{\eV}{\,{\rm eV}}

\newcommand{\omegap}{\omega_{\rm pl}}
\newcommand{\etal}{\emph{et al. } }
\newcommand{\Pphi}{P_{\gamma \leftrightarrow \phi}}
\newcommand{\be}{\begin{eqnarray}}
\newcommand{\ee}{\end{eqnarray}}
\newcommand{\ba}{\left( \begin{array}{cc}}
\newcommand{\ea} {\end{array} \right)}
\newcommand{\bv}{\left( \begin{array}{c}}
\newcommand{\ev} {\end{array} \right)}

\begin{document}
\bibliographystyle{unsrt}

\title{Active Galactic Nuclei Shed Light on Axion-like-Particles}
\author{Clare Burrage}
\affiliation{Theory Group, Deutsches Elektronen-Synchrotron DESY,
  D-22603 Hamburg, Germany}
\author{Anne-Christine Davis}
\affiliation{Department of Applied Mathematics and Theoretical Physics,
Centre for Mathematical Sciences,  Cambridge CB3 0WA, United Kingdom}
\author{Douglas J. Shaw}
\affiliation{Queen Mary University of London, Astronomy Unit, School of Mathematical Sciences, Mile End Road, London E1 4NS, United Kingdom}
\date{\today}
\begin{abstract}
We demonstrate that the scatter in the luminosity relations of astrophysical objects can be used
to search for axion-like-particles (ALPs).  This analysis is applied to
observations of active galactic nuclei, where we find evidence highly suggestive of the existence of a very light ALP.
\end{abstract}
\preprint{DESY 09-023}
\maketitle
Precision observations of the recent Universe are an invaluable test-bed for `new physics' such as the existence of new, low mass and/or weakly interacting particles. Such a particle, the axion, was proposed in 1977 by Pecci and Quinn to solve  the strong CP problem \cite{CP}.  Since then, other models which also feature 
(very) light, neutral spin zero, axion-like-particles (ALPs) have been proposed e.g. \cite{AxionModels, Brax07, AxionPDG}. 
Recently, analyses of starlight polarization \cite{Burrage08}, and the distribution and spectrum of high energy cosmic rays \cite{AxionHints} have provided tentative evidence for ALPs. 

In this \emph{Letter},  we present a new method for studying ALPs using astrophysical X/$\gamma$-ray luminosity relations, which, when applied to observations of Active Galactic  Nuclei (AGN), provides the strongest evidence yet for the existence of ALPs.

ALPs are scalar or pseudo-scalar
particles, $\phi$, which couple to photons, $A^{\mu}$, via  the terms
$(\phi/M)F_{\mu\nu}F^{\mu \nu}$ or $(\phi/M)\epsilon_{\mu
  \nu\rho\sigma}F^{\mu \nu}F^{\rho \sigma}$   in the Lagrangian
respectively; $F_{\mu \nu} = 2\partial_{[\mu}A_{\nu]}$. 
 We define $m_{\phi}$ to be the ALP mass, and $g_{\gamma\gamma\phi} = 1/M$ 
is the coupling between ALPs and photons. 

In the presence of a background magnetic field, $B$, ALPs mix with
photons. The probability that an ALP converts into a photon whilst
travelling through a coherent  magnetic domain of length $L$ is \cite{Raffelt88}
\be
\Pphi = \sin^2 2\theta \sin^2 \left(\frac{\Delta}{\cos 2\theta}\right),
\ee
where $\Delta = m_{\rm eff}^2 L/4\omega$, $\tan 2\theta = 2B \omega
/ M m_{\rm eff}^2$, $m_{\rm eff}^2 = m_{\phi}^2 - \omegap^2 -
\epsilon B^2/M^2$; $\hbar$, $c=1$.  $\omegap^2 =4\pi \alpha_{\rm em} n_{\rm
  e}/m_{\rm e}$ is the plasma frequency; $n_{\rm e}$ is the
electron number density, $m_{e}$ the electron mass and $\alpha_{\rm
  em} \approx 1/137$ is the electromagnetic fine structure
constant. $\epsilon = +1$ for scalars and $0$ for pseudo-scalars;
generally $B^2/M^2 \Vert m_{\phi} - \omegap^2\Vert \ll 1$. The total flux (in
ALPs and photons) is conserved by the mixing process, however photon
number is not. 

ALP-photon mixing is constrained by a number of laboratory experiments
(see Ref. \cite{AxionPDG} and references therein) but the
tightest constraints come from the astrophysical consequences of
ALPs (see Ref. \cite{Raffelt08}).

We are concerned with very light ALPs,  $m_{\phi} \ll 10^{-12}\eV$.
For such masses, observations of the supernova SN1987A limit $g_{11} =
10^{11}\GeV/M \lesssim 1$ \cite{AxionPDG} for pseudo-scalars, whilst limits on
new long ranges forces require $g_{11} < 10^{-16}$ for scalars. 
However, if ALPs are chameleonic, these constraints do not apply \cite{Brax07,Burrage08}, and the best constraint comes from the structure of  starlight polarization: $g_{11} \lesssim 100$ \cite{Burrage08}. 

As photon number is not conserved ALP-photon mixing in the magnetic
fields of galaxies and galaxy clusters alters the observed luminosity
of astrophysical objects.  
In this \emph{Letter}, we consider light that has travelled through $N \gg 1$ randomly oriented magnetic regions.  
This is true of light from many astrophysical sources, particularly
those in galaxy clusters. We focus on the limit in which  the mixing
is strong, $N\Pphi \gg 1$, and frequency independent, $N\Delta
\lesssim \pi/2$.

In this strong mixing limit, little or no circular polarization  is
produced \cite{Burrage08} and power is spread with equal probability
between the two photon polarization states $\gamma_{1}$
and $\gamma_{2}$ and $\phi$.  We take  $I^{\rm
  (tot)\,1/2}\mathbf{u} =(\gamma_{1},\gamma_{2}, \phi)^{T}$ to
be real, then since $\Vert \mathbf{u}\Vert^2 = 1$ is conserved  $\mathbf{u}$ describes a point on a sphere, and any point is as likely as any other. Thus, after strong mixing, the normalized final state $\mathbf{u}$ is a random vector with:
\be
\mathbf{u} = \left(\sqrt{1-K^2}\cos \pi\Theta,  \sqrt{1-K^2}\sin \pi\Theta,  K \right)^{T}, \nonumber
\ee
where $K, \Theta \sim U[-1,1)$, and  the final flux in the photon
field is  $I_{\rm f}^{(\gamma)} = (1-K^2)I^{\rm (tot)}$. The initial photon flux is $I^{(\gamma)}_{0}$ and we assume $I^{(\gamma)}_{0} \approx I^{\rm (tot)}$.  A state
labelled by a real vector $\mathbf{u}$ is sufficient to describe any
mixture of $\phi$ with fully linearly polarized light. We extend this
result to light with partial or no initial linear polarization 
by noting that any such state, with $I^{(\gamma)}_{0} \approx I^{\rm (tot)}$, can be written as a superposition of two
real $\mathbf{u}$ state vectors \cite{Burrage08}.  Where $p_{0}$ is
the initial degree  of linear polarization, the final photon flux after strong mixing is:
\be
I_{\rm f}^{(\gamma)}(K_1,K_2,p_{0}) = \left[1 - (1-p_0)K^2_1 - p_0 K^2_2\right]I^{\rm (tot)},
\ee
where $K_{i} \sim U(-1,1)$. We define $C \equiv I_{\rm
  f}^{(\gamma)}/I^{\rm (tot)}$. Averaging $C$ over many different
light paths through a large number of   randomly oriented magnetic
regions gives $\bar{C} = 2/3$. This average reduction in the apparent
luminosity of astrophysical sources is well known \cite{Csaki02}, however
attempts to use it to constrain ALPs have, so far, been unsuccessful
since the intrinsic luminosity of the astrophysical objects is not
known with sufficient accuracy. It is rarely appreciated that $C=2/3$ only when averaged over many different light paths (e.g. light from many different objects). For a given light path through the magnetic regions, there is significant scatter and skew of $C$ about its mean of $2/3$ \cite{Burrage08}.  The probability that $C \in [c-\dd c,c+\dd c]$ is $f_C(c)\dd c$ where:
\be
f_{C}(c;p_0) &=& \frac{1}{\sqrt{1-p_{0}^2}}\left[ \tan^{-1}\left(\sqrt{a}\left(1-\frac{2c_{+}}{1+p_0}\right)^{-1/2}\right)\right. \nonumber\\ &&- \left. \tan^{-1}\left(\sqrt{a}\left(1-\frac{2c_{-}}{1-p_0}\right)^{1/2}\right)\right] 
\label{scatter}
\ee
where $a = (1-p_0)/(1+p_0)$ and $c_{\pm} = \min(c, (1\pm p_0)/2$.

The central idea of this \emph{Letter} is that the scatter in
empirically established luminosity relations can constrain,
detect or rule out strong mixing between ALPs and photons. 
Provided $\omega$ is high enough, strong and (almost) frequency independent mixing can be caused by the magnetic fields of galaxy clusters \cite{Burrage08}, the existence of which is well established \cite{Carilli02}. When such mixing occurs, 
there will be a contribution to the scatter of the observed luminosities as described by Eq. (\ref{scatter}).  
Observations  imply that most clusters contain 
magnetic fields of strength $B\approx 1-10 \mu{\rm G}$ which are generally coherent over 
$L \sim {\rm kpc}$ scales, although $L$ may be as large as $10-100\,{\rm kpc}$ in some cases 
\cite{Carilli02}.  Typical electron densities in the diffuse plasma in the intra-cluster
 medium (ICM) are $n_{e} \sim 10^{-3}\,{\rm cm}^{-3}$ and hence $\omegap \sim
10^{-12}\eV$, and $m_{\phi}^2\ll \omegap^2$ if
$g_{11}\sim\mathcal{O}(1)$.  This last condition requires the ALP coupling to be very close to 
the upper bound from supernovae or for the ALP to be chameleonic in which case $g_{11} \gtrsim 1$ is allowed.

 If light travels a typical distance 
of $0.1-1{\rm Mpc}$ through the ICM then $N \approx 100-1000$ magnetic domains will have been traversed. 
Strong mixing requires $N\Pphi \gg 1$, 
so $\sqrt{N}(BL/2M) \gg 1$ and 
$\omega \gg M\omegap^2 / 2\sqrt{N} B$.  With $B_{5} = B/5\mu G$, 
$L_{2} = L/2\kpc$ and  $g_{11} = 10^{11}\GeV/M$ we have $BL/2M \approx 0.15 g_{11} B_{5} L_{2}$.  
Typically $B_{5}, L_{2} \sim \mathcal{O}(1)$ and for $N \sim 100
-1000$ and we need $\omega \gg 16 - 51\eV$. Frequency independent mixing, $N\Delta \lesssim \pi/2$, i.e. 
requires $\omega \gtrsim  N(2\pi/\omegap^2L) = 3-30\,{\rm keV}$. Numerical simulations 
show that the frequency independent limit is still approximately valid for frequecies that are a factor of $10$ smaller i.e. 
$\omega\gtrsim 0.3-3\,{\rm keV}$.  When $N\Pphi \gg 1$, $N\Delta \gg 1$ the measured 
luminosity is attenuated by a factor of $2/3$ with 
relatively little scatter.

For the remainder of this \emph{Letter}, we assume that $\omega \gtrsim 2\keV$ light, i.e. X/$\gamma$ rays, which originated in or passed through a galaxy cluster, has undergone strong and (almost) frequency
independent ALP  mixing, requiring $g_{11} \gtrsim 0.1 -0.3$ and
$m_{\phi} \lesssim 10^{-12}\eV$.  For $\omega \ll 0.5\keV$, mixing
is highly frequency dependent and can be either weak (so $C \approx 1$) or strong. In the latter case the luminosity is reduced by a factor $\approx 2/3$. 

 We require that the X/$\gamma$ sources are compact i.e. their size, $R$, is $\lesssim L \sim {\rm few} \,{\rm
  kpc}$. Diffuse light, such as the  X-ray light from galaxy clusters ($R \sim 10^2 -10^{3} \kpc$) is  not suitable for our analysis as it  will have traced many different paths through the magnetic fields of the cluster, and the effects of mixing will be
averaged over all of these paths.  For such sources, the only effect of strong mixing is the 2/3
suppression of the total luminosity.  

For a number of classes of compact astrophysical objects, correlations
between the  X or $\gamma$ ray luminosity or radiated energy and some
feature  of their light-curve (e.g. peak energy) or the object's
luminosity at a lower frequency have been empirical established. We
let $Y_{i}$ label the X or $\gamma$ ray luminosity / total energy and $X_{i}$ label the light-curve feature or lower energy luminosity with which it is correlated.  The relations  between $Y_{i}$ and $X_{i}$ take the form:
\be
\log_{10} Y_{i} = a + b\log_{10} X_{i} + S_{i}. \label{corrForm}
\ee
where $S_{i}$ vanish on average, and represent the scatter of individual measurements about the mean relation. The scatter comes partly from measurement error but in most cases the largest contribution appears to be intrinsic (e.g. \cite{Schaefer07}).  It is standard practice to model the $S_{i}$ as being normally distributed with mean $0$ and some variance $\sigma^2$ i.e. $S_{i} = \sigma \delta_{i}$ where $\delta_{i} \sim N(0,1)$.  We refer to this as the Gaussian scatter model.  If the high frequency light has been subject to strong mixing with an ALP we expect:
\be
S_{i} = \sigma \delta_{i} - \log_{10} C_{i} + \mu, \label{ASMmodel}
\ee
and  $C_{i}$ has p.d.f. $f_{C}(c)$ as given above. $\mu$ is the
expectation of $\log_{10} C_{i}$, so the $S_{i}$ still have mean $0$;
$\mu$ can always be absorbed into a redefinition of the fitting
parameter $a$. We call this the ALP strong mixing (ALPsm) scatter model. 
 The distribution of the $\log_{10} C_{i}$ is   both a distinct feature of
strong mixing and very different from a normal distribution.  Provided the variance of the intrinsic Gaussian scatter ($\sigma^2$ in both models) is not too large, it is possible, with enough measurements,
to use the  distribution of the scatter to constrain, detect or rule out such strong mixing.
We do this by means of a likelihood ratio test, comparing the null Gaussian hypothesis with the ALPsm hypothesis. Both models have the form $S_{i} = \sigma \delta_{i} + \log_{10}((1-f) + fC_{i})$; $f$ parametrizes the fraction of light that is strongly mixed. $0 < f< 1$ corresponds to partial strong mixing. However, along
a given path, either strong mixing occurs or it does not; the X or $\gamma$ ray light from an object cannot be
partially strongly mixed.   $f$ is not therefore  a free parameter to
be fitted.  The likelihood, $L_{f}(a,b,\sigma,p_{0})$ of the model with general $f$ is:
\be
L_{f}(a,b,\sigma, p_{0}) = \prod_{i}\frac{1}{\sqrt{2\pi}\sigma}\int_{0}^{1} e^{-\frac{z_{i}^2}{2\sigma^2}} f_{C}(c;p_{0}) \dd c.
\ee
where $z_{i} =\log_{10}Y_{i}-a -b\log_{10} X_{i} - h(c;f)$ and $h(c;f)
= \log_{10}((1-f)+f c)$.  We fix $f$ and $p_{0}$  and  find the values
of $a=\hat{a}$, $b=\hat{b}$ and $\sigma=\hat{\sigma}$ which maximize $L_{f}$. We define
$\hat{L}_{f}(p_0) = L_{f}(\hat{a},\hat{b},\hat{\sigma}, p_{0})$, and
calculate 
$$r_{f}(p_{0}) = 2\log \left(\hat{L}_{f}(p_0)/\hat{L}_{0}\right).$$
Keeping $p_{0}$ fixed, both the Gaussian and ALPsm models have the same number of free parameters.  This means
that $r_{1}(p_{0})$ is equivalent to the Bayesian
Information Criteria  commonly used for model selection.  Conventionally, $r_{1}(p_0) <
-6 \, (r_{0} >  6)$ would be `strong evidence' against (for) the ALP strong mixing model over the Gaussian one. $\Vert r_{1} \Vert > 10$ corresponds to `very strong evidence'. If ALPs are preferred, a useful check is to ensure that $r_{f}(p_0)$ is maximized for $f \approx 1$.   If this is not the case, we would conclude that whilst the data is not compatible with simple Gaussian scatter, it is also not particularly indicative of the strong mixing with ALPs.

Luminosity relations of the required form exist for Gamma Ray Bursts
(GRBs) \cite{Schaefer07}, Blazars \cite{Bloom07, Xie97} and  AGN \cite{Steffen06}
and are suitable for our analysis.  Additionally an $\mathcal{O}(1)$ fraction of such
 objects are expected to be located within galaxy clusters.

The $\gamma$-ray luminosity, $L_{\gamma}$, and radiated energy,
$E_{\gamma}$, of  Gamma Ray Bursts (GRBs) have been found to be
correlated with a number of spectral features, giving $5$ seemingly independent relations (see \cite{Schaefer07}).
Additionally the $L_{\gamma}$ of Blazars, a class of AGN, is correlated with both their radio wave, $L_{\rm r}$  \cite{Bloom07} and near infra-red (IR) \cite{Xie97} luminosities, $L_{\rm k}$. We analyzed observations of 69 GRBs Ref. \cite{Schaefer07}, with redshifts $z = 0.17 - 6.6$, 95 EGRET observations of Blazars with  $z \approx 0.02-2.5$ Ref. \cite{Bloom07} for the radio relation, and 16 Blazars ($z \approx 0.3 - 1$) for the IR relation \cite{Xie97}. For all the relations however, either data points were too few or 
the intrinsic Gaussian scatter was too large to constrain ALP mixing. For $4$ out of $5$ GRB relations and both Blazar relations, $r_{1} > 0$ but in all cases $\vert r_{1} \vert <
0.75$; the sum of these $r_{1}$'s is only $r_{1}^{\rm GRB, \,Blazar} \approx 1.6$ for $p_{0}=0$ (with similar values for other $p_{0}$)  a statistically insignificant preference for the ALPsm model.

There is also a strong correlation between the  $2\keV$ monochromatic X-ray
luminosity, $L_{\rm X}$, of AGN and their monochromatic optical
luminosity, $L_{\rm o}$ (at $2500$\AA; $\omega \approx 4.95\eV$)
\cite{Steffen06}. This relation is of the form: $\log L_{X} \approx a
+b \log L_{o}$.  We use observations of 77 optically selected AGNs
with  $z=0.061-2.54$ from the COMBO-17 and ROSAT surveys as
tabulated in Ref. \cite{Steffen06} to analyse the scatter in this
relationship. For $0 < p_{0} < 0.4$ we find:
\be
r_{1}(p_0)^{\rm AGN} \approx 14,
\ee
and for all $p_0$, $r_{1}(p_0)^{\rm AGN} > 11$.  Additionally
$r_{f}(p_{0})$ is strongly peaked at $f=1$: $r_{1}(0) \approx 14.1$,
$r_{0.95}(0) \approx 8.6$ and $r_{0.9}(0) \approx 4.9$. 

There is clearly a structure in the scatter fitted better by ALP mixing
 than by the Gaussian scatter model.   It is not clear, however,
 whether this is due to the success of the former or the failure of
 the latter which was only  adopted for convenience, and because for other relations (e.g those of GRB or Blazars), it provides a good fit to the scatter.  It may be that if AGN physics were better understood, a null hypothesis for the scatter distribution would be predicted that is a better fit than the ALP mixing model.   Whilst we cannot rule out this scenario we can,  independent of any null hypothesis, qualitatively check whether the structure of the scatter is really well matched by the ALP mixing model.  
\begin{figure}[tbh]
\begin{center}
\includegraphics[width=72mm]{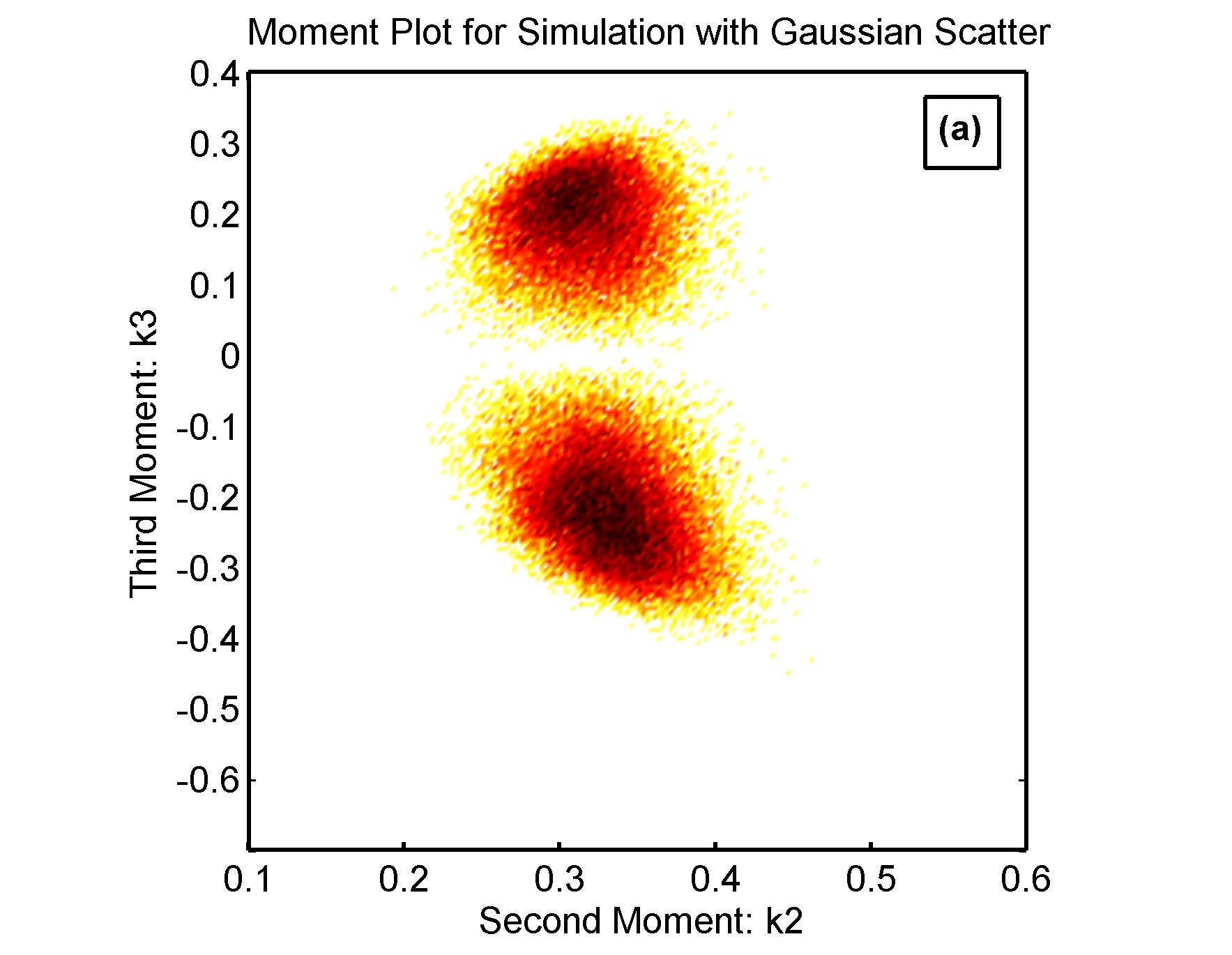} 
\includegraphics[width=72mm]{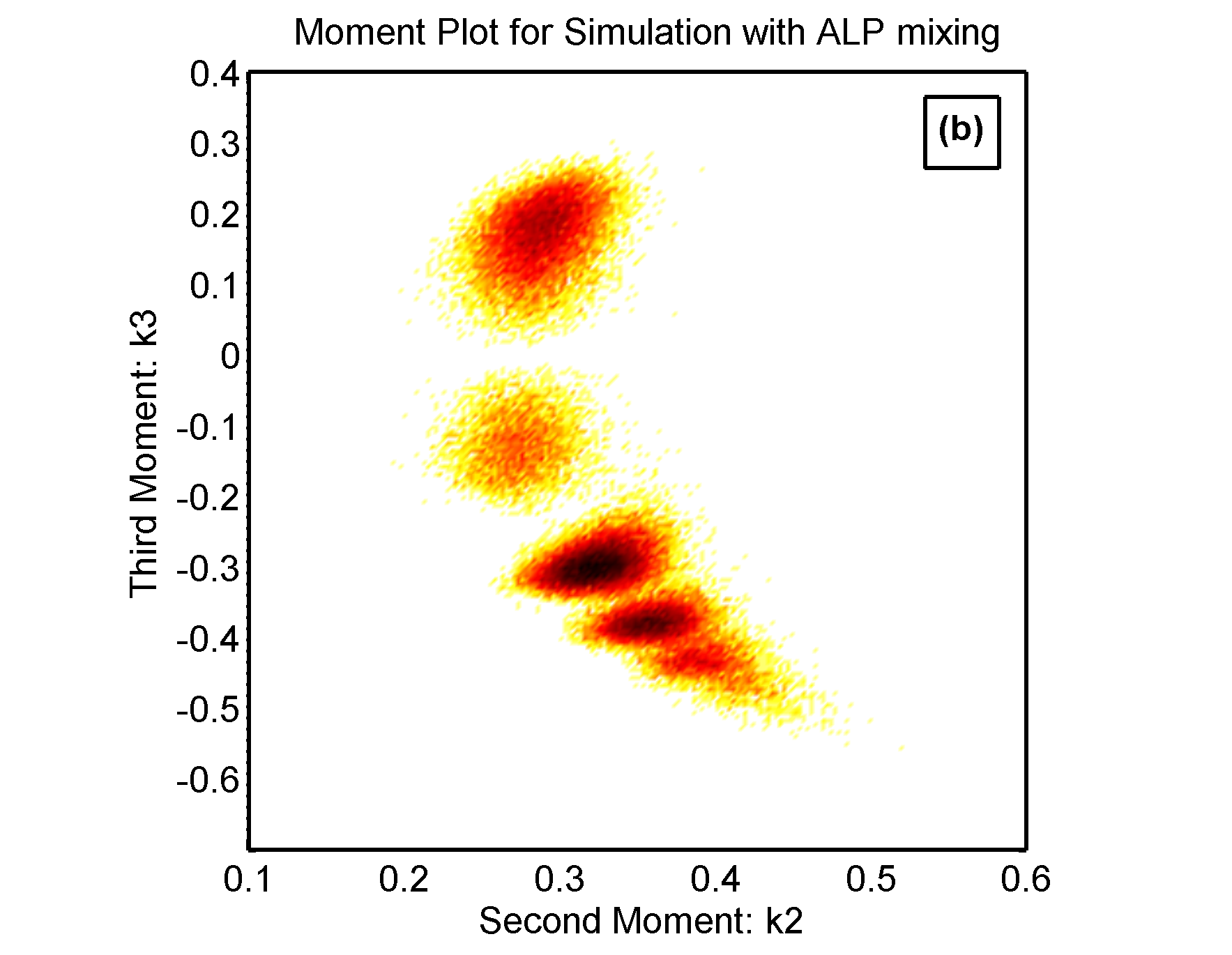}
\includegraphics[width=72mm]{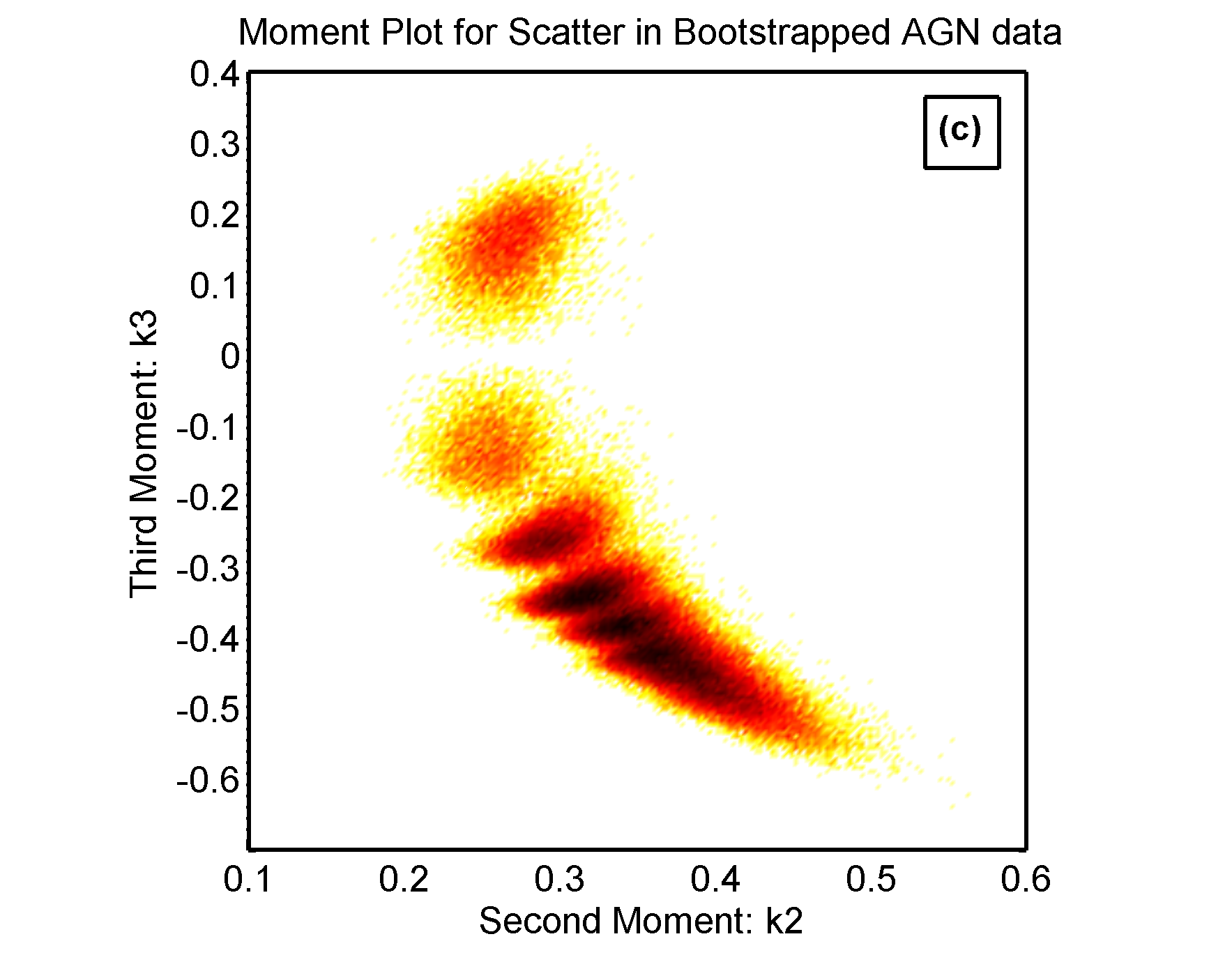}
\end{center}
\caption{(Colour online) Histograms of second ($k_2$) and third ($k_{3}$) moments (see Eq. \ref{momentDef})   for $10^{5}$ resamplings of 77 data points (a) simulated with the best-fit Gaussian scatter ($\sigma = 0.34$) (b) simulated with the best-fit ALP mixing model ($\sigma = 0.23$) and (c) from the observed scatter in the AGN $L_{\rm X}-L_{\rm o}$ relation. Darker regions indicate higher density. See text for discussion.
\label{fig1}}
\end{figure}

We perform such a check by making $10^5$ bootstrap re-samplings with replacement, 
$D^{\ast} = \left \langle x_{i}^{\ast},y_{i}^{\ast}\right \rangle$, of the original data, 
$D = \left \langle x_{i}\equiv \log_{10} X_{i},y_{i} \equiv \log_{10} Y_{i}\right \rangle$.
Both $d$ and the $D^{\ast}$ have 77 data points. For each $D^{\ast}$ we match the relation 
$s^{\ast}_{i} = y_{i}^{\ast} - (a^{\ast} + b^{\ast}x_{i}^{\ast}) = 0$ by minimizing the $\sigma_{\rm RMS} = k_2$ where
\be
k_{m}(\left\lbrace s_{i}\right\rbrace) = \left(\frac{1}{N_{p}}\sum_{i} s_{i}^{\ast m}\right)^{\frac{1}{m}}, \label{momentDef}
\ee
where $N_{p}$ is the number of data points. $k_{2}$ is the RMS average of the $s_{i}$ and 
$k_{3}^3/k_{2}^{3}$ is their skewness.  2d histograms of $k_{i}$ vs. $k_{j}$ reveal non-trivial
 correlations between the $k_{m}$ and, unlike the likelihood analysis, are relatively insensitive to
any outlying data points. For comparison, we simulate data sets for
both the best-fit Gaussian and ALPsm models in which $\hat{\sigma} = 0.34$ and 
$\hat{\sigma} = 0.23$ respectively. We ensure that for each original simulated 
data set $k_{2} \approx 0.34$ and plot $\left\lbrace k_{2},k_{3}\right\rbrace$ for the $10^5$ resamplings.

The detailed form of the plots varies from simulation to simulation, however a number of qualitative 
features can be identified as `fingerprints' of the Gaussian or ALPsm models.
In the former, there are  two  density peaks around 
$\left \lbrace k_{2},k_{3}\right\rbrace \approx \left \lbrace 0.34, \pm 0.25\right\rbrace$, 
whereas in the latter similar peaks often occur around $\approx \left\lbrace 0.23 - 0.3,  \pm 0.15\right\rbrace$.
The main fingerprint of the best-fit ALPsm model, which occurs in most, if not all, of the 
simulations, is that most of the data points fall in a long `tail' 
$\left \lbrace k_{2},k_{3}\right\rbrace\sim \left\lbrace 0.3, -0.3\right\rbrace$ to 
$\left\lbrace (0.4 - 0.5),  -(0.5 - 0.7)\right\rbrace$.  These features persist for different values
 and distributions of $p_{0} \lesssim 0.5$, and in more realistic simulations where only a fraction ($\gtrsim 50\%$)
the objects have been subject to strong ALP mixing. These features can be clearly seen in FIGs \ref{fig1}a \& b which are respectively typical $k_{2}-k_{3}$ histograms for data sets simulated with the best-fit Gaussian and ALPsm models. Darker regions indicate higher density.
 
FIG. \ref{fig1}c is the $k_{2}-k_{3}$ plot for the actual AGN data. There is a marked qualitative 
similarity between FIG. \ref{fig1}c  and the sample ALPsm model plot, FIG. \ref{fig1}b. 
A dense tail-like feature is clearly present and its extent, direction and structure of peaks 
is typical of that seen in the ALP simulations.  Although not shown here, there is also a strong
 similarity between the AGN $k_{3}-k_{5}$ plot and those found in the best-fit ALPsm model simulations. 
No evidence for any  correlation between redshift and 
scatter was found, ruling out an explanation for it based on evolution  of the $L_{\rm X}-L_{\rm o}$ 
relation and / or an inaccurate choice of cosmological model.  We note
that this strong mixing could be
independently verified by a measurement the polarization of $2\keV$
light from the AGN \cite{Burrage08}.

In this \emph{Letter}, we have shown that the scatter in empirical
X/$\gamma$ ray luminosity relations can be used to constrain mixing
between ALPs and photons. When applied to the AGN $L_{\rm
  X}-L_{\rm o}$ relation, this shows strong evidence for ALPs relative to the null hypothesis of Gaussian scatter.  Additionally, the visualizations of the AGN data reveal a scatter distribution with a strong qualitative similarity to that predicted by the best-fit ALP-photon strong mixing model. This similarity is independent 
of the null hypothesis. Strong mixing of ALPs with ${\rm keV}$
photons will take place in galaxy clusters if $M \lesssim {\rm few} \times 10^{11}\GeV$ and $m_{\phi} \ll 10^{-12}\eV$, or in magnetic fields close to the AGN if $M \sim 10^{10} \GeV$ and $m_{\phi} \ll 10^{-7}\eV$ \cite{AGNmag}.  Whilst we cannot rule out explanations of the scatter in terms of 
known physics, it is, at the very least, a  remarkable coincidence
that both this and other recent analyses \cite{Burrage08, AxionHints}
are fitted better by models in which very light ALPs (with similar  couplings and masses) exist.

\acknowledgments CB acknowledges the German Science
Foundation (DFG) under the Collaborative Research Centre (SFB) 676. ACD \& DJS acknowledge STFC.



\end{document}